\documentclass{aa} 
\usepackage{graphicx}
\usepackage[sectionbib,authoryear]{natbib} 
\usepackage{times}
\usepackage{graphicx} % when using Latex and dvips
\def\simge{\mathrel{\rlap{\raise 0.511ex \hbox{$>$}}{\lower 0.511ex 
  \hbox{$\sim$}}}}
\def\simle{\mathrel{\rlap{\raise 0.511ex \hbox{$<$}}{\lower 0.511ex 
  \hbox{$\sim$}}}}
\newcommand{\gsim}{\,\raisebox{0.2em}{$>$}\!\!\!\!\!
\raisebox{-0.25em}{$\sim$}\,}
\newcommand{\lsim}{\,\raisebox{0.2em}{$<$}\!\!\!\!\!
\raisebox{-0.25em}{$\sim$}\,}
\newcommand{\gr}{$\gamma$-ray}
\newcommand{\grs}{$\gamma$-rays}
\newcommand{\rxj}{RX~J0852.0-4622}
\newcommand{\hess}{H.E.S.S.}

\begin{document}

\title{Theory of cosmic ray and \gr\ production in the supernova remnant \rxj}

%   \subtitle{}

\authorrunning{Berezhko et al.}  
\titlerunning{CR and \gr\ production in SNR \rxj} 

\author{ E.G. Berezhko \inst{1}
         \and G. P\"uhlhofer \inst{2,}\thanks{now at 
Institut f\"ur Astronomie und Astrophysik, Universit\"at T\"ubingen, Sand 1, 72076 T\"ubingen, Germany }
          \and H.J. V\"olk \inst{3}
}

\institute{Yu.G. Shafer Institute of Cosmophysical Research and Aeronomy,
                     31 Lenin Ave., 677980 Yakutsk, Russia\\
              \email{berezhko@ikfia.ysn.ru}
	      \and
	 Landessternwarte, K\"onigstuhl, D-69117 Heidelberg, Germany\\
	   \email{ Gerd.Puehlhofer@lsw.uni-heidelberg.de}
         \and
           Max-Planck-Institut f\"ur Kernphysik,
                Postfach 103980, D-69029 Heidelberg, Germany\\
              \email{Heinrich.Voelk@mpi-hd.mpg.de}           
}
	     
\offprints{H.J. V\"olk}

\date{Received month day, year; accepted month day, year}

\abstract {} 
{The properties of the Galactic supernova remnant (SNR) \rxj\ are
  theoretically analysed.}  
{An explicitly time-dependent, nonlinear kinetic
  model of cosmic ray (CR) acceleration in SNRs is used to describe the
  properties of SNR \rxj, the accelerated CRs and the nonthermal emission. The
  source is assumed to be at a distance of $\approx 1$~kpc in the wind bubble
  of a massive progenitor star. An estimate of the thermal X-ray flux in such a
  configuration is given.}  
{We find that the overall synchrotron spectrum of \rxj\ as well as the
  filamentary structures in hard X-rays lead to an amplified magnetic field $B
  > 100 \mu$G~ in the SNR interior. This implies that the leptonic very high
  energy (VHE) \gr\ emission is suppressed, and that the VHE \grs\ are
  hadronically dominated. The energy spectrum of protons produced over the
  life-time of the remnant until now may well reach ``knee'' energies. The
  derived \gr\ morphology is consistent with the H.E.S.S. measurements. The
  amount of energy in energetic particles corresponds to about 35\% of the
  hydrodynamic explosion energy.  A remaining uncertainty concerns the thermal
  X-ray flux at 1~keV. A rough estimate, possibly not quite appropriate for the
  assumed wind bubble configuration, results in it being larger than the
  nonthermal flux at this energy.}
{It is concluded that this SNR expanding into the wind bubble of a massive star
  in a dense gas environment can be a hadronic \gr\ source that is consistent
  with all existing multi-wavelength constraints, except possibly the thermal
  X-ray emission.}

\keywords{(ISM:)cosmic rays -- +acceleration of particles -- shock waves --
supernovae individual (SNR \rxj) -- radiation mechanisms:non-thermal --
gamma-rays:theory}

\maketitle

%
%________________________________________________________________

\section{Introduction}

\rxj\ (also known as G266.2-1.9) is a shell-type supernova remnant (SNR) with a
diameter of $2^{\circ}$, located in the Galactic plane. The SNR was originally
discovered in X-rays with the ROSAT satellite
\citep{aschen98,aschenbach1999rosat}. In projection along the line of sight,
\rxj\ lies entirely within the still much larger Vela SNR and is only visible
in hard X-rays, where the thermal radiation from the Vela SNR is no longer
dominant. While nonthermal emission from the shell of \rxj\ has been confirmed
by several X-ray observatories \citep{slane01,bamba05,iyudin2005xmm}, a
clear detection of the thermal X-ray emission from the shell or the interior
was not yet possible because of confusion with the Vela SNR. This implies that
the thermal emission is very weak\footnote{Very recently \citet{uchiyama08}
has indicated that the X-ray emission from \rxj\ might show a thermal
component.}.

The radio emission of \rxj\ is also weak. In fact the SNR radio shell was only
identified \citep{combi1999radio,duncang00,stupar2005radio} after its discovery
in X-rays. Before that only a bright radio spot was known as ``Vela Z''
\citep{milne1968velasnr}, which was usually identified with the Vela SNR. In
addition, significant Galactic background variation over the size of the
remnant cannot be excluded. The radio spectrum of \rxj\ is therefore not well
determined. Only for the northeastern rim a spectral index can be derived with
moderate accuracy \citep{duncang00}.

The shell of \rxj\ was also detected in very high energy (VHE) \grs\ by the
H.E.S.S. collaboration \citep{aha05,aha07a}, with a \gr\ flux at 1 TeV as large
as that from the Crab Nebula. Emission from the northwestern rim had been
detected already before by the CANGAROO experiment
\citep{cangaroo2005velajr}. The CANGAROO data have been revised since then
\citep{enomoto06}.

\rxj\ is the second SNR after RX~J1713.7-3946 \citep[e.g.][]{aha07b}, where
morphologically a SNR shell was unambiguously identified to accelerate
particles to TeV energies and beyond. As a third and fourth source of this
character the objects RC W86 \citep{hlg07,aha08} and SN 1006 \citep{naumann09}
have recently been detected. A few other TeV sources have been detected in the
\hess\ Galactic plane survey that are spatially coincident with radio
shell-type SNRs. So far, however, the data do not permit to unambiguously
identify the \gr\ emission with the SNR shells, e.g. using morphological
arguments. In several cases, the TeV emission might also be associated with
X-ray emitting pulsar wind nebula candidates.

Despite the scarcity of precise information from radio data and despite
uncertainties about key astrophysical parameters, the prominence of \rxj\ has
therefore prompted us to model the acceleration of both electrons and protons
together with their nonthermal emission in detail, applying explicitly
time-dependent nonlinear kinetic theory. The theory couples particle
acceleration on a kinetic level with the gas dynamical evolution of the system
\citep{byk96,bv97,bv00}.

Compared to the other SNRs that were successfully described within the
framework of this theory \citep[e.g.][]{ber05,ber08,bv06}, the present
uncertainties regarding \rxj\ are quite large. Such important astronomical
parameters as the distance and age are not well known. It is in fact not even
clear, whether the source is in front or behind the Vela SNR. The latter object
is generally considered to lie at a distance $d=250 \pm 30$~pc
\citep{cha1999velasnrdistance}\footnote{The distance to the Vela pulsar is
  estimated at $d=287^{+19}_{-17}\mathrm{pc}$ \citep{dodson03}, see also
  \citet{caraveo01}.}. A position behind the Vela SNR could correspond to a
solution with $d=1$~kpc \citep{slane01}, whereas another solution could
correspond to the earlier distance estimate of $d=200$~pc \citep{aschen98}.

  This prompted us originally to construct indeed two quite different source
  scenarios, one in front, and the other behind the Vela SNR. They were to
  correspond to earlier distance estimates: a ``nearby '' solution with
  $d=200$~pc \citep{aschen98} in front of, and a ``distant'' solution with
  $d=1$~kpc \citep{slane01} behind the Vela SNR.  However, no ``nearby''
  solution could be found that fulfilled all the observational
  constraints. Therefore we have abandoned the possibility of a small distance
  to \rxj\ and will assume for the sequel that the source is at a distance
  $d=1$~kpc.

  In agreement with \citet{slane01} but with independent additional arguments
  we shall assume that the observed nonthermal emission of \rxj\ indicates that
  the SNR emerged from a core collapse explosion into the wind bubble of a
  massive progenitor star in a dense gas environment, possibly a
  molecular cloud. In this case the major part of the swept-up volume has
  originally been occupied by the highly diluted bubble gas that also has a
  minimal thermal emissivity. At the current epoch, however, we assume that the
  SNR shock already propagates into the dense shell of ambient {\it
    interstellar} medium (ISM) which has originally been compressed by the
  stellar wind. This also implies that the magnetic field upstream of the SNR
  shock is of interstellar origin.

  We note here that this solution has a similar character as the solution
  suggested earlier for the object RX~J1713.7-3946 \citep{bv06}. Indeed, in
  this sense the two SNRs \rxj\ and RX~J1713.7-3946 can be considered as twins.

  The hydrodynamic state of the system, i.e. the given linear radius for known
  angular radius and distance together with the present age and shock velocity,
  etc., is basically determined by the choice of mechanical explosion energy,
  ejected mass, and external density.

The lack of knowledge of the radio spectral shape makes it impossible to derive
from the radio synchrotron observations -- and as far as the magnetic field is
concerned, from the combined radio and X-ray synchrotron spectral observations
-- the most relevant pair of physical parameters for the acceleration theory,
namely the proton injection rate and the effective magnetic field strength
\citep{bkv02,vlk04,ber05,ber08}. Thus, even if we consider the circumstellar
medium (CSM) structure to be given, it is not possible to directly {\it
predict} the form of the overall synchrotron and the full VHE \gr\ emission
from theory.

The observed overall nonthermal spectral shape -- including the VHE \grs\ --
and the small-scale filamentary structures in the nonthermal X-ray emission of
\rxj\ nevertheless provide evidence for effective nuclear cosmic ray (CR)
acceleration, associated with considerable magnetic field amplification. This
conclusion is possible because the overall nonthermal spectrum can be
theoretically {\it fitted} with an appropriate proton injection rate,
electron-to-proton ratio, and effective magnetic field strength (assumed to be
uniform inside the shocked CSM cf. \citet{bv04b}).

The main result of this paper is that the resulting VHE \gr\ flux is
  hadronically dominated. The well-known difference in the effectiveness of the
  basic radiation mechanisms then implies that the energy density of the
  nuclear component of the nonthermal charged-particle population in the SNR by
  far dominates that of the energetic electron component generated in the
  source. The energy in nonthermal particles at the present epoch amounts to
  $\sim 35$~percent of the assumed total mechanical energy $E_\mathrm{sn}=
  1.3\times 10^{51}$~erg, released in the explosion. Therefore, from the point
  of view of energetics, this solution for \rxj\ more than fulfils the
  average requirement on a SNR source of the Galactic CRs. In this context we
  shall also qualitatively discuss the question of possible escape of the
  highest-energy particles, accelerated at an earlier phase of the SNR than the
  one we can observe at present.

  In the next section the theoretical model is described. Section 3 presents
  our assumptions regarding the values of the physical parameters as they are
  suggested by the broadband data and by physical considerations. It contains
  also a discussion of the thermal emission, even though
    it has not been possible to estimate it appropriately for the assumed wind
    bubble and shell configuration. The results for the
  gas dynamics, the particle acceleration coupling with it, the \gr\ emission,
  and for the thermal X-ray emission are presented and discussed in section
  4. In section 5 our conclusions are summarized.

\section{Model}

The theoretical model for the particle acceleration combined with the gas
dynamics of the explosion has been described earlier, for instance in a recent
analysis of SNR RX~J1713.7-3946 \citep[see][and references therein]{bv06}.

The ejected mass $M_\mathrm{ej}$ has initially in its fastest moving outer
parts a power law distribution $dM_\mathrm{ej}/du\propto u^{2-k}$ in flow
velocity $u$, with $7<k<12$ \citep{Jones81,chev82}. We shall choose $k=8$ (as
in the case of SNR RX~J1713.7-3946).

The interaction of the ejecta with the CSM creates a strong outward-propagating
shock wave in the CSM at which particles are accelerated. Our nonlinear model
is based on an explicitly time-dependent solution of the CR transport equations
together with the gas dynamic equations in spherical symmetry. In particular
the theory takes into account the adiabatic energy losses of thermal and
nonthermal particles in the SNR interior, the diffusion of nonthermal particles
from the outer shock into that interior, and the backreaction of CRs on the
shock structure and dynamics. This backreaction decelerates the thermal gas
already in front of the shock and leads to a smooth shock precursor that
reduces the Mach number of the subsequent collisionless plasma shock (the
subshock), heating the inflowing gas\footnote{It is implicitely assumed here
that any energy loss due to radiative gas cooling has no effect on the shock
structure. Given the rather high internal temperature $T_\mathrm{sub}$~of the
remnant at this stage (see section 4.2) this appears to be a safe assumption.}.

However, nuclear particles are only effectively injected into the acceleration
process at those parts of the moving shock surface, where the locally mean
magnetic field vector is quasi-parallel to the shock normal. These regions are
characterized by magnetic flux tubes within which the injection of moderately
suprathermal particles proceeds.

In quasi-perpendicular shock regions, on the other hand, injection is
instantaneously strongly suppressed or completely prohibited. In the extreme
there are then separated regions at the shock surface, where particle injection
is permanently either allowed or prohibited. Then also no nuclear particles are
accelerated in the prohibited regions. An example for this extreme situation is
SN 1006, where the X-ray emitting polar cap regions correspond to the
quasi-parallel regions. In these allowed regions diffusive shock acceleration
strongly proceeds on the Bohm diffusion level (see below) as a result of the
effective injection of low-energy particles. However, since Bohm diffusion is
isotropic, the energetic particles can also cross field lines and thus possibly
reach quasi-perpendicular regions of the shock where they can also
accelerate. The extent to which this happens depends on the spatial scales that
separate quasi-perpendicular regions from quasi-parallel regions. For a
homogeneous external magnetic field about 80\% of the shock surface is
quasi-perpendicular in the above sense, and therefore only a fraction
$f_{\mathrm {re}} \approx 0.2$ of the shock is efficiently accelerating
\citep{vbk03}. However we shall argue in section 3.1.4 that for a SNR,
propagating into a stellar wind bubble with a radiatively cooling shell of
high-density gas, the above spatial scales are probably so small that
$f_{\mathrm {re}} \approx 1$. This has substantial implications 
especially for the resulting thermal X-ray emission, because the
  observed VHE \gr\ emission (which is by implication of hadronic origin) then
  requires a lower gas density.

For given magnetic field strength the electron injection rate in spherical
symmetry is determined by the intensity of the observed overall synchrotron
spectrum. For given injection rate of nuclear particles and magnetic field
strength the ratio $K_\mathrm{ep}$ between the densities of nonthermal
electrons and nuclear particle can then be calculated.

As a result of the streaming instability the accelerating CRs very effectively
excite large-amplitude magnetic fluctuations upstream of the SN shock
\citep{bell78,bo78,mv82}. Since these fluctuations scatter CRs
extremely strongly, we approximate the CR diffusion coefficient $\kappa (p)$ by
its lower limit, corresponding to a scattering mean free path equal to the
particle gyro radius. In this so-called Bohm limit $\kappa (p)= [mc^3/(3eB)]
(v/c)(p/mc)$, where $e$ and $m$ are the particle charge and mass, $v$ and $p$
denote the particle velocity and momentum, respectively, $B$ is the effective
magnetic field strength (see below), and $c$ is the speed of light. Regarding
the nuclear particles with the highest energies this Bohm limit may imply an
underestimate for $\kappa (p)$ also for another reason, since for these
particles the effective, amplified field has spatial scales that are
smaller or equal to their gyro radius \citep{bell04,pell06}. To this extent our
assumption of Bohm diffusion in the amplified field yields an upper limit to
the maximum energy of the nuclear particle \citep[e.g.][]{ZirakPtus2008}. In
addition we assume that the interior effective magnetic field has a uniform
strength after its MHD compression in the thermal subshock. Practically
speaking we assume this uniformity over a spatial scale that is large compared
to the thickness of the observed X-ray filaments \citep{bv04b}. For a different
point of view, see \citet{pohl05}.

Another important nonlinear effect of the strong excitation of magnetic
fluctuations by the accelerating particles themselves is the heating of the
thermal plasma in the shock precursor that is generated by the accelerating
particles. Combined analytical and numerical efforts, using plasma theory to
give a nonlinear description of the magnetic field evolution
\citet{lucb00,belll,PtuskinZirak2003,bell04,pell06}, concluded that a
considerable amplification of the upstream magnetic field should occur in the
acceleration process to what we call the effective magnetic field. The physical
reason is that the beam of efficiently accelerated {\it nuclear CR component}
excites in particular also a non-resonant magnetic mode in addition to the
well-known resonant Alfv{\'e}n waves \citep{bell04}. The latter have
nevertheless been argued to contribute dominantly to the overall turbulent
magnetic energy density in the shock precursor \citep{pell06}. However, the
three-dimensional MHD simulations of the non-resonant instability by
\citet{bell04} and \citet{Ziraketal2008} show that the nonlinear growth of the
magnetic fluctuations is accompanied by the formation of internal shocks and
correspondingly strong dissipation which heats the thermal
plasma\footnote{Particle-in-cell simulations by \citet{np08} led these authors
  even to the extreme conclusion that the non-resonant instability amplitudes
  become never large as a result of the bulk acceleration of the thermal gas by
  the streaming CRs. We believe that in a quasi-steady shock configuration the
  CR current is being steadily driven through the upstream gas by the
  downstream overpressure, as implied above.}.  Analogous dissipation should
occur in three dimensions for the wave modes of the resonant streaming
instability. We approximate this physical process by assuming that the heat
input into the thermal gas equals the (linear) growth of the turbulent field
energy in the excited Alfv{\'e}n waves {\it in the already amplified effective
  field} \citep[see][for the correspondence of this approach to existing theory
and experiment]{ber08,vbk08}.

As already mentioned in the Introduction, without a reliable spectral
index for the observed spatially-integrated radio synchrotron emission we shall
choose the strength of the effective field as well as 
the nuclear injection rate $\eta$, such as to optimally
fit the calculated synchrotron spectrum (from radio to X-ray energies) 
together with the calculated VHE \gr\ spectrum to the
observations. Subsequently this spectrally fitted magnetic field is
compared with the field derived from the observed filamentary
structure. The degree of agreement between these two field strengths is then
used as a measure of the success and self-consistency of the model.

The filament-based magnetic field strength $B_\mathrm{d}$ downstream of the
shock is determined by the observed width $L$ of the X-ray filament --
interpreted as the synchrotron cooling length behind the shock -- through the
relation
\begin{equation}
B_\mathrm{d}=[3m_e^2c^4/(4er_0^2l_2^2)]^{1/3}(\sqrt{1+\delta^2}-\delta)^{-2/3},
\label{eq1}
\end{equation}
where $\delta^2=0.12c/(r_0\nu)[V_\mathrm{s}/(\sigma c)]^2$, 
$l_2\approx L/7$ is the
radial width of the X-ray emissivity $q_{\nu}(\epsilon_{\nu},r)$,
$\epsilon_{\nu}$ is the X-ray energy, corresponding to the frequency $\nu$, and
$r_0$ denotes the classical electron radius \citep{bv04a}. This field
strength is clearly a lower limit.

Since we have already used the observed amplitude of the VHE \gr\ spectrum to
determine $\eta$ we do not predict this amplitude, as one could do using a
detailed knowledge of the integrated synchrotron spectrum. The above
consistency condition for the effective field still needs to be fulfilled for
the solution to be acceptable.

Moreover, we shall not only investigate whether the chosen values for $B$ and
$\eta$ are consistent with the {\it Chandra} measurement of the X-ray
filamentary structure \citep{bamba05}, but also to which extent they are
consistent with the semi-empirical relation \citep{bv06}
\begin{equation}
B_0^2/(8\pi P_\mathrm{c}) \approx 5\times 10^{-3}
\label{eq2}
\end{equation}
which connects the upstream magnetic field pressure $B_0^2/(8\pi)$ in the shock
precursor (i.e. upstream, but amplified by the CR instability) with the
pressure $P_\mathrm{c}$ of the accelerated particles that drives the field
amplification in the first place. Eq.(2) holds for a number of SNRs that could
be analyzed with the aid of a well-known radio spectrum for these sources
\citep{vbk05a}. The degree of fulfilment of this relation is a further quality
criterium for the model to judge its success in theoretically describing the
TeV \gr\ source. Eq.(2), or a relation of a similar type \citep{belll}, is
likely to hold for an individual object also during its time evolution. Yet, in
order to avoid the introduction of a further theoretical parameter, we
shall consider $B_0$ as constant in time, equal to its present value, in our
evaluation of the models for \rxj. We shall come back to this point in section
4.3.

In this specific form the three theory ``parameters'' $B$, $\eta$, and
$K_\mathrm{ep}$ are determined {\it quantitatively} by comparison with the
observations at the present age of the source. Their time-dependence during the
evolution of the SNR is disregarded in this paper.  

The numerical solution of the dynamical equations at each instant of time
yields the CR spectrum and the spatial distributions of CRs and thermal
gas. This allows the calculation of the spectra of the expected fluxes of
nonthermal emission produced by the accelerated CRs, the morphology of the
emission, and the future evolution inasmuch the same physical processes
continue to work at later times.

In the following we shall consider the wind bubble scenario for \rxj\ in
the general framework of this model.

\section{Physical parameters of \rxj}

\begin{table}[ht]
%\vspace{-1.5pc}
\caption{Key model parameters, and corresponding spectral,
  dynamical, and morphological values expected from the calculations.} 
\label{T:modelparameters}
\begin{center}
  \begin{tabular}{l|l}
\hline 
%& Vela Jr  \\ 
$d$ & 1\,kpc \\ 
$R_\mathrm{s}$ & 17.5\,pc \\
\hline \\ 
$B_{\mathrm{d}}$ from X-ray filaments & 139$\,\mathrm{\mu G}$ \\  
\hline
$E_{\mathrm{sn}}$ & $1.3 \times 10^{51}\mathrm{erg}$ \\ 
$M_{\mathrm{ej}}$ & $3.5\,M_{\odot}$ \\ 
$N_{\mathrm{g}}(R_{\mathrm{s}})$ & $0.24\,\mathrm{cm^{-3}}$ \\ 
$N_{\mathrm{g}}(r=0)$ & $0.003\,\mathrm{cm^{-3}}$ \\ 
$k$ & 8 \\ 
$B_{0}$ & $20\,\mathrm{\mu G}$ \\ 
$\eta$ & $3 \times 10^{-4}$ \\ 
$K_{\mathrm{ep}}$  & $3 \times 10^{-4}$ \\ 
$f_{\mathrm{re}}$ & 1  \\ 
\hline \\ 
$t_{\mathrm{sn}}$ & 3745\,yr   \\ 
$V_{\mathrm{s}}(t_{\mathrm{sn}})$ & $1316\,\mathrm{km\,s^{-1}}$ \\ 
$\sigma(t_{\mathrm{sn}})$ & 5.2 \\ 
$\sigma_{\mathrm{s}}(t_{\mathrm{sn}})$ & 3.1  \\ 
$M_\mathrm{s}(t_{\mathrm{sn}})$ & $25~M_{\odot}$ \\
$E_{\mathrm{c}}(t_{\mathrm{sn}})$ & $4.6 \times 10^{50}\mathrm{erg}$ \\ 
$B_{\mathrm{d}}(t_{\mathrm{sn}}) (=\sigma B_{0})$ & $104\,\mathrm{\mu G}$ \\ 
\hline
\\ $P_{\mathrm{c}}/(\rho_{0}V_{\mathrm{s}}^{2})$ & 0.145 \\ 
$B_{0}^{2}/(8\pi P_{\mathrm{c}}) $ & $ 6.5\times 10^{-3}$ \\ 
\hline
\hline
  \end{tabular}
\end{center}
%\vspace{-2pc}
%Key model parameters for the solution, and corresponding spectral,
%dynamical and morphological values expected from the calculations. 
Parameter description: The quantities $d$ and $R_\mathrm{s}$ denote the assumed distance and the radius
  of the source, respectively, $B_\mathrm{d}$ is the internal magnetic field
  strength, as determined from the thickness of observed X-ray filaments
  cf. eq. 1, and $E_\mathrm{sn}$ is the total hydrodynamic explosion energy;
  $M_\mathrm{ej}$, $M_\mathrm{s}(t_{\mathrm{sn}})$,
  $N_\mathrm{g}(R_\mathrm{s})$ and $N_\mathrm{g}(r=0)$ are the ejected mass,
  the swept-up mass, the circumstellar gas number density at the SNR shock, and
  the number density at the centre, respectively; $k$ is the power law index of
  the ejecta velocity distribution; $B_0$ is the assumed amplified magnetic
  field strength in the upstream region of the shock precursor, while $\eta$
  and $K_\mathrm{ep}$ denote the assumed proton injection rate and energetic
  electron-to-proton ratio, respectively; $t_\mathrm{sn}$ is the calculated age
  of the SNR; $V_\mathrm{s}(t_\mathrm{sn})$, $\sigma(t_\mathrm{sn})$,
  $\sigma_\mathrm{s}(t_\mathrm{sn})$, $E_\mathrm{c}(t_\mathrm{sn})$, and
  $B_\mathrm{d}(t_\mathrm{sn})$ are the resulting values of the subshock
  velocity, the total compression ratio, the subshock compression ratio, the
  total nonthermal energy, and the downstream magnetic field strength,
  respectively. Finally, $P_\mathrm{c}$ and $\rho_0 =
  m_\mathrm{p}N_\mathrm{g}(R_\mathrm{s})$ denote the postshock pressure of
  accelerated particles and postshock mass density of the gas, respectively.
\end{table}
%\vspace{-1.5pc}

%\newpage

In this section, we will describe the physical parameters of the model. Section
3.1 gives an overview over the available broadband data. Section 3.2 describes
the setup of the scenario. Because of the scarcity -- and sometimes ambiguity
-- of the available data, we will use some general arguments from the
non-thermal particle emission not only to constrain the acceleration parameters
but also the environmental parameter gas density and its spatial distribution.
The values of all relevant physical parameters are given in Table 1.

\subsection{Broadband data of \rxj}
\label{SS:broadband}

\subsubsection{Morphology of \rxj\ in general}

\rxj\ is a shell-type SNR with a shell diameter of $2^{\circ}$, as seen in hard
X-rays \citep{aschen98,aschenbach1999rosat,slane01} and VHE
$\gamma$-rays \citep{aha05,aha07a}. Also the radio continuum emission
correlates spatially well with the high energy emission
\citep{duncang00,stupar2005radio,aha07a}. This confirms the shell-type
nature of the source, although confusion with the emission from the Vela SNR
prevented a detection based on the radio data alone.  

In X-rays, a central diffuse source of $\sim 9' \times 14'$ extension is seen
NW of the geometrical centre of the SNR
\citep{slane01,becker2002NSwithXMM,becker2007NSwithXMM}.  While \citet{slane01}
argue that the hard spectrum of the source seen with ASCA hints at a Pulsar
Wind Nebula (PWN), \citet{becker2007NSwithXMM} argue, based on XMM-Newton data,
that the source is soft and reject a PWN nature. Hence, there is currently no
agreement whether \rxj\ is a centre-filled, composite SNR or
not. Spatially-integrated fluxes would only marginally be affected. There is
however the possibility that such a central PWN would slightly influence the
VHE $\gamma$-ray radial profile \citep[see][and
Fig.\,\ref{F:gammaradial}]{aha07a}.

\subsubsection{Central compact object and pulsar association}

At the geometrical centre of the shell of \rxj\ lies the X-ray point source
CXOU\,J085201.4-461753
\citep{aschenbach1999rosat,pavlov2001central,kargaltsev2002central,
mereghetti2001central,mereghetti2002central}, a central compact object (CCO)
similar to that detected in the centre of Cas\,A.  CXOU\,J085201.4-461753 might
be a neutron star, but the nature of the object and the possibly associated
compact H$\alpha$ nebula \citep{pellizzoni2002central} is still under debate
\citep[e.g.][]{reynoso2006radiopointsources,mignani2007vlacco,
becker2007NSwithXMM}. No X-ray pulsations have been detected
\citep{kargaltsev2002central,becker2007NSwithXMM}. The association of
CXOU\,J085201.4-461753 with \rxj\ is nevertheless suggestive and, if true,
would allow conclusions on the nature of the progenitor of \rxj. Since the
scenario discussed in this paper implies a core collapse SN, it does not
exclude an association of CXOU\,J085201.4-461753 with \rxj.  

In the literature also the possibility is discussed that \rxj\ is associated
with PSR\,J0855-4644 \citep{redman2005pulsar}. This appears unlikely however,
given the implications of this association on distance and age of \rxj\
\citep{redman2005pulsar}.

\subsubsection{Non-thermal X-rays}

The soft X-ray emission from \rxj\ is heavily confused by thermal emission from
the Vela SNR. While the emission from the Vela SNR seems consistently
constrained to two thermal components ($T_{1,2} = 0.05, 1.2\,\mathrm{keV}$)
with an absorption column density of about $10^{20}\mathrm{cm}^{-2}$
\citep{lu2000vela}, the surface brightness and temperatures of these components
are variable enough to prevent a clean subtraction of the Vela SNR in the X-ray
spectra of \rxj\ \citep{slane01,iyudin2005xmm,aha07a}. In the spectra above
$\sim 1\,\mathrm{keV}$, the non-thermal emission from \rxj\ nevertheless
clearly dominates. We adopt the common interpretation that this component is
synchrotron emission from relativistic electrons and use two estimates of the
total X-ray synchrotron flux from \rxj\ in our modelling (see
Figs.\,\ref{F:broadband}, \ref{F:synchrotron}): The first
(lower) estimate was derived using the averaged power-law spectra derived from
the three brightest shell components \citep{slane01}, which we scaled up to
match the total flux measured with ROSAT in the soft X-ray band
\citep{aschen98}. The second (upper) estimate is a reanalysis of the ASCA data
set as presented in \citet{aha07a}.

High-resolution imaging of this synchrotron X-ray component with XMM-Newton
\citep{iyudin2005xmm,iyudin2007xmm} and especially with Chandra
\citep{bamba05,pannuti2004chandra} permits the derivation of synchrotron
cooling times and therefore of an estimate of the effective magnetic field
\citep[e.g.][]{bkv03,bv04a,bamba05}. In order to test our model we will compare
the B-field derived in this manner to the field value that is required to fit
the spatially integrated synchrotron data (see section 4).

\subsubsection{Thermal X-rays}
The detection of thermal X-ray emission could help to constrain the gas density
and/or its spatial distribution in the SNR.  However, the interpretation
of the X-ray spectra is impeded by the strong background emission from the Vela
SNR; the latter is dominated by low-temperature ($< 1$~keV) X-ray
emission. After subtraction of such emission the remaining emission (dominant
at higher energies) has been attributed either to gas at higher temperatures
from \rxj\ \citep{aschenbach1999rosat}, or to synchrotron emission plus a very
small thermal contribution \citep{slane01}. The latter conclusion has basically
been taken over in the recent {\it H.E.S.S.} paper \citep{aha07a} that also
contains a re-analysis of the ASCA data, even though the X-ray line features
exhibited in the spectrum below 2 keV might be associated with either \rxj\ or
the Vela SNR, or to both. Similarly the residuals of the spectra to the ASCA
data visible around 1 keV are suggestive that another component, which might
originate from \rxj, is needed. The possible detection of a thermal
  emission component from \rxj\ by \citet{uchiyama08} is to be mentioned here
  again.

These are complex possibilities. The calculation of the thermal emission for
the concrete case of \rxj\ is compounded by the wind-bubble plus swept-up-shell
geometry. Standard methods for calculating the thermal emission from SNRs
approximate the configuration either by a plane shock geometry or by a
classical Sedov solution, and they neglect the existence of the accelerated
particle component. Neither of these approximations is appropriate for the case
of \rxj\ in its present phase. The plane approximation disregards the adiabatic
gas cooling in the interior and the overall dynamic evolution of the
system. The classical Sedov solution implies a uniform circumstellar
medium. In the extreme case the radiatively cooled wind bubble shell is even
denser -- and therefore even thinner -- than assumed in subsection 3.2.2. and
it might only be reached recently before the present epoch. Then collisional
electron heating has had little time to operate until now.

Another important aspect is the modification of the SNR shock by the
accelerating CRs. As mentioned before, for a SNR shock propagating into a
uniform medium with a uniform magnetic field this implies that the larger part
of the shock surface corresponds to a quasi-perpendicular shock with a strongly
reduced injection of nuclear particles \citep{vbk03}\footnote{For earlier
  discussions of this question in a more general context, see
  \citet{ellison95,malkov95}.}. Suprathermal injection of ions is only possible
in the quasi-parallel shock regions. If the spatial scales of the
quasi-perpendicular regions are large enough, then the cross-field diffusion of
the highest-energy particles, accelerated in the magnetic flux tubes
delineating the neighboring quasi-parallel shock regions, does not reach deeply
into these quasi-perpendicular regions. In the corresponding magnetic flux
tubes there are no nuclear particles to be accelerated, there is no magnetic
field amplification, and the shock remains unmodified there. This means that in
the quasi-perpendicular regions the shock dissipation and therefore the gas
heating occurs in a locally unmodified shock with the overall shock speed,
leading to a correspondingly high gas temperature and high thermal emission. In
the case where a radiatively cooling shell of a wind bubble is the obstacle for
the SNR expansion, the situation may be different. MHD instabilities and the
radiative cooling of such a shell probably break it into many small regions
with strongly varying field directions. Then the spatial scales separating the
flux tubes originating from quasi-perpendicular regions from those of
quasi-parallel shock regions may become small enough that cross-field
diffusion can also fill the quasi-perpendicular flux tubes with accelerating
particles and then particle acceleration plus magnetic field amplification
occur practically everywhere over the shock surface \citep{vbk08,voelk08}. In
the extreme this implies shock modification over the entire shock region and
thus a reduced gas heating due to the subshock dissipation only. The enhanced
overall acceleration efficiency then also demands a lower density of the
thermal gas for a given hadronic \gr\ flux, and thus a lower thermal
emission. The low swept-up mass in a low-density wind bubble in addition lowers
the overall thermal emissivity compared to that of a classical Sedov remnant
with the same upstream gas density at the shock at the present epoch.

In section 4.4 a rough estimate of the resulting thermal X-ray emission will be
given, based on the emission from a classical Sedov solution in a uniform
ambient medium. According to this estimate the thermal emission of soft X-rays
at 1 keV is larger than the corresponding nonthermal X-ray emission. However,
the error in this estimate is not known and is likely to be quite
large. Therefore an uncertainty remains which we cannot resolve at this point.

\subsubsection{X-ray morphology absorption and relation to CO data}

Absorption of soft X-rays by neutral hydrogen can be used to put constraints on
the source distance.  \citet{slane01} used CO data of the Vela Molecular Ridge
(VMR) to reject a distance of \rxj\ of more than 1-2\,kpc, based on the lack of
strong X-ray absorption variation that should have been detected across
\rxj. This is broadly in agreement with the inference by
  \citet{moriguchi2001velasnrinco} that there is an anticorrelation of X-ray
  emission and molecular gas traced in CO, especially with regard to the VMR at
  a distance of $1-2$~kpc.

In the $2-10\,\mathrm{keV}$ band, that is expected not to exhibit absorption,
an X-ray power-law spectrum can be derived. Assuming this power-law to continue
down to $0.7\,\mathrm{keV}$, \citet{slane01} have then derived an absorption
column density of $(4.0 \pm 1.8) \cdot 10^{21}$~cm$^{-2}$ for \rxj.  Since this
column density is much larger than the one towards the Vela SNR, they concluded
that \rxj\ should be at a much larger distance than the Vela SNR.

\subsubsection{Radio spectrum}

We assume that the radio emission is due to synchrotron radiation.  There is no
good radio spectrum available fvelaj61.dvior the entire remnant.  We use the 
differential flux values given by \citet{duncang00} at 2.42\,GHz and
1.40\,GHz, the errors of which are representing the uncertainty of the
background level.  The spectral index between the two bands has quite a large
error ($\alpha = 0.4 \pm 0.5$), but for the north-western rim a better spectrum
($\alpha = 0.40 \pm 0.15$) could be derived.  If this value is representative
for the entire remnant, as \citet{duncang00} argue, then this index is somewhat
harder than what we expect from a modified SNR shock environment, though still
compatible within a $2\,\sigma$ error range.  

\subsubsection{Gamma-ray and X-ray line emission from radioactive 
$^{44}$Ti decay}

The $^{44}\mathrm{Ti}$ production in a SN explosion depends on progenitor star
mass and explosion-type, with a yield spanning two orders of magnitude
\citep[see, e.g.][and references therein]{Renaud06}.  Because of the short
lifetime of $\sim 80$ years \citep[e.g.][]{wietfeldt1999ti44lifetime}, the mere
detection of hard X-ray and $\gamma$-ray de-excitation lines at 69.7, 78.4, and
1157\,keV from the $^{44}\mathrm{Ti}$ radioactive decay products can be used to
significantly constrain the SNR age. So far, however, these lines have only
been detected unambiguously from Cassiopeia A (age presumably $\sim 330$
years), with COMPTEL onboard CGRO \citep{iyudin1994comptelcasa}, PDS onboard
{\it BeppoSAX} \citep{vink2001saxcasa}, and the ISGRI imager onboard INTEGRAL
\citep{renaud2006ti44casa}.

For \rxj, the situation is unfortunately unresolved.  From COMPTEL data, the
detection of a $\gamma$-ray line at $1163\pm16\,\mathrm{keV}$, consistent with
the $^{44}\mathrm{Ti}$ $\gamma$-ray decay line, was claimed and predominantly
attributed to \rxj\ \citep{iyudin1998comptel,aschenbach1999rosat}.  Using this
$^{44}\mathrm{Ti}$ line flux, and using a rather high shock velocity,
\citet{aschenbach1999rosat} derived an age of $680\pm 100$~yrs, and a distance
of $200$~pc. Such \gr\ data also suggest a core collapse SN event.

However, \citet{schoenfelder2000ti44} pointed out that the significance of the
COMPTEL $^{44}\mathrm{Ti}$ result is only marginal.  And the COMPTEL result
could so far not be confirmed with the INTEGRAL instruments SPI and ISGRI.  A
SPI upper limit is so far unconstraining \citep{vonkienlin2005integral}.  Under
a point-source assumption, the ISGRI non-detection of the 78\,keV line would be
in conflict with the COMPTEL result \citep{Renaud06}. However, an
extended-source analysis has not yet been performed
\citep{renaud2006extendedintegral}.

Therefore a nearby and rather recent event is not decisively excluded from
these specific observations, even though the arguments for it are rather
weak.

\subsection{Model parameters for a SN explosion in 1\,kpc distance}

As discussed in the previous section, the main observational motivation to
locate \rxj\ at a distance of $\sim 1\,\mathrm{kpc}$ is the much larger column
density in neutral hydrogen derived from the X-ray spectrum of \rxj, compared
to the values for Vela SNR.  

\subsubsection{Reasons for a wind bubble scenario}

The lack (or low level) of thermal X-ray emission and the strong X-ray
synchrotron flux already led \citet{slane01} to the conclusion
that \rxj\ could be evolving into a wind bubble. Similarly,
\citet{duncang00} argue that the unusual radio properties of \rxj\
(bipolar shell morphology, low surface brightness) could be explained if the
SNR has so far evolved in a low-density region.  

We can use the non-thermal X-ray plus $\gamma$-ray emission from \rxj\ to put
these arguments for a wind bubble scenario on a more firm footing,
anticipating for the moment that the VHE $\gamma$-ray emission is indeed
dominated by hadronic emission. As we shall see in section 4, this latter
conclusion follows from the fact that with the assumption of a significantly
amplified magnetic field it is possible to fit all the nonthermal spectra as
well as the morphology in nonthermal X-rays and gamma-rays -- and that this
amplified field does not exceed that derived from the observed X-ray
filaments. Such amplification is however only possible by the nonthermal
pressure of the nuclear particles.

Arguments similar to those given below have been used in \citet{bv06}
for the comparable case of SNR RX\,J1713.7-3946 to which we refer the reader
here.

In order to yield the observed nonthermal X-ray luminosity in the case of
 a uniform ISM the shock speed should be sufficiently large, $V_s \gsim
10^3$~km/s \citep{bv04b}. Given that the SNR would already be in the Sedov
phase the observed size $R_\mathrm{s} \sim 20\,\mathrm{pc}$, corresponding to a
distance of 1 kpc, would lead to the age constraint $t_\mathrm{sn}\approx
0.4R_\mathrm{s}/V_\mathrm{s}\lsim 3\times 10^3$~yr. For a typical SN type Ia
explosion energy $E_\mathrm{sn}=10^{51}$~erg and ejected mass of $1.4
M_{\odot}$ this would then imply a very low ISM number density
$N_\mathrm{H}\lsim 10^{-2}$~cm$^{-3}$. On the other hand, the peak TeV \gr\
luminosity, achieved during SNR evolution from a type Ia event, roughly scales
as \citep{bv97}:
\begin{equation}
\epsilon_{\gamma}F_{\gamma}(\epsilon_{\gamma})\approx
150 \left(\frac{N_\mathrm{H}}{1\mbox{~cm}^{-3}}\right)
\left(\frac{1\mbox{~kpc}}{d}\right)^2
~~~\frac{\mbox{eV}}{\mbox{cm}^2\mbox{s}},
\end{equation}
for $f_\mathrm{re}= 1$. Here
$F_{\gamma}(\epsilon_{\gamma})$ is the integral flux of \grs\ with energies
greater than $ \epsilon_{\gamma}$. This expression shows that for
$N_\mathrm{H}<10^{-2}$~cm$^{-3}$ we would have to expect an energy flux
$\epsilon_{\gamma}F_{\gamma}(\epsilon_{\gamma})<1.5$~eV/(cm$^2$s). This is
an order of magnitude smaller than the observed flux (see section 4).

Therefore the nonthermal observations make it clear that SNR \rxj\ can not
correspond to a type Ia event, if the source distance is as large as
$d=1$~kpc. As a consequence we shall consider a core collapse SN event.

\subsubsection{Wind bubble parameters}

The progenitor stars of core collapse SNe that significantly modify the density
of their environment are massive main-sequence stars with initial masses
$M_\mathrm{i}>15M_{\odot}$ which have intense winds, e.g.  \citep{abb}. In the
mean, during their evolution in the surrounding uniform ISM of gas number
density $\rho_0=m_\mathrm{p} N_\mathrm{ISM}$, they create a low-density bubble,
surrounded by a shell of swept-up and compressed ISM of radius
\citep{weaver,chevl}
\begin{equation}
R_\mathrm{sh}=0.76( 0.5 \dot{M}V_\mathrm{w}^2 t_\mathrm{w}^3/\rho_0)^{1/5},
\label{eq20}
\end{equation}
where $\dot{M}$ is the mass-loss rate of the progenitor star, $V_\mathrm{w}$ is
the wind speed, and $t_\mathrm{w}$ is the duration of the wind phase. The
values of these parameters are given in tabular form by \citet{chev82} in terms
of $M_\mathrm{i}$.

In order to determine the SNR shock dynamics inside the shell we model the gas
number density distribution in the bubble and in the shell in the form
\citep[e.g.][]{bv06}:
\begin{equation}
N_\mathrm{g}=N_\mathrm{b}+(r/R_\mathrm{sh})^{3(\sigma_\mathrm{sh}
-1)}N_\mathrm{sh},
\label{eq21}
\end{equation}
where $N_\mathrm{sh}=\sigma_\mathrm{sh} N_\mathrm{ISM}$ is the peak number
density 
{\tiny per hydrogen atom} 
in the shell, $N_\mathrm{b}$ is the gas number
density inside the bubble, typically very small compared with the shell
density, and $\sigma_\mathrm{sh} = N_\mathrm{sh}/N_\mathrm{ISM}$ is the shell
compression ratio. We note that as a result of radiative cooling the
compression ratio $\sigma_\mathrm{sh}$ can exceed the classical upper limit of
4. The possibility of a very thin wind bubble with $\sigma_\mathrm{sh}\gg
4$ has been indicated in subsection 3.1.4. above.

The mass of the bubble 
\begin{equation}
M_\mathrm{b}=(4\pi R_\mathrm{sh}^3/3)m_\mathrm{p} N_\mathrm{b}
\label{eq22}
\end{equation}
is rather small, $M_\mathrm{b}< M_{\odot}$, in the case of moderate progenitor
masses $M_\mathrm{i}<20M_{\odot}$, whereas the shell mass
\begin{equation}
M_\mathrm{sh}=4\pi N_\mathrm{sh}m_\mathrm{p}\int_0^{R_\mathrm{sh}}dr
r^2(r/R_\mathrm{sh})^{3(\sigma -1)}= (4\pi
R_\mathrm{sh}^3/3)N_\mathrm{ISM}m_\mathrm{p}
\label{eq23}
\end{equation}
is many hundred solar masses. Therefore, during
SNR shock propagation through the bubble, only a small fraction of its energy
is given to gas of stellar origin. The main part of the explosion energy is
deposited in the shell.

Here we use the gas number density distribution
$N_\mathrm{g}(r)=\rho(r)/m_\mathrm{p}$ in the form
\begin{equation}
N_\mathrm{g}=\{0.003+0.24[r/(17.5~\mbox{pc})]^{12}\}~~\mbox{cm}^{-3},
\end{equation}
which fixes the gas density at
the present shock radius and provides a consistent fit for all existing data
for SNR \rxj. Together with Eq.(4), this relation also connects the external
density $N_\mathrm{ISM}$ with the progenitor mass.

Such a distribution corresponds to a bubble with $\sigma_\mathrm{sh}=5$ and
$28<R_\mathrm{sh}<32$~pc created by the wind of a main-sequence star of initial
mass $15M_{\odot}<M_\mathrm{i}<20M_{\odot}$ in the surrounding ISM of hydrogen
number density $11<N_\mathrm{ISM}<49$~cm$^{-3}$, respectively \citep{chevl}. It
implies that this bubble is located inside a region of dense gas. 

\subsubsection{Further parameters that determine the SN evolution and CR 
 acceleration}

We shall use the SNR parameters $E_\mathrm{sn}=1.3\times 10^{51}$~erg,
$M_\mathrm{ej}=3.5M_{\odot}$, and $k=8$ which, as will be shown below, give a
good fit for the observed SNR properties.

We shall also use an upstream effective magnetic field value $B_0=20$~$\mu$G,
which is required to provide the observed synchrotron flux in the radio and
X-ray bands (see below). Such a value of $B_0$ is significantly higher than a
merely MHD-compressed dense-gas magnetic field in the inner part of the shell,
at densities $N_\mathrm{g} \leq 0.24~\mbox{cm}^{-3}$.

\section{Results and discussion}

In this section we shall discuss the physical characteristics of the wind
bubble scenario in detail, and compare them with the observations. 

The calculated dynamical characteristics of the SNR are shown in Fig.\,1. From
Fig.\,1a one can see that for the assumed distance of 1 kpc the calculation
fits the observed SNR size $R_\mathrm{s}\approx 17.5$~pc at the age
$t_{\mathrm{sn}} = 3745$~yr.

%-----------------------------------------------------------------------fig.1
\begin{figure}
\centering
\includegraphics[width=8.cm]{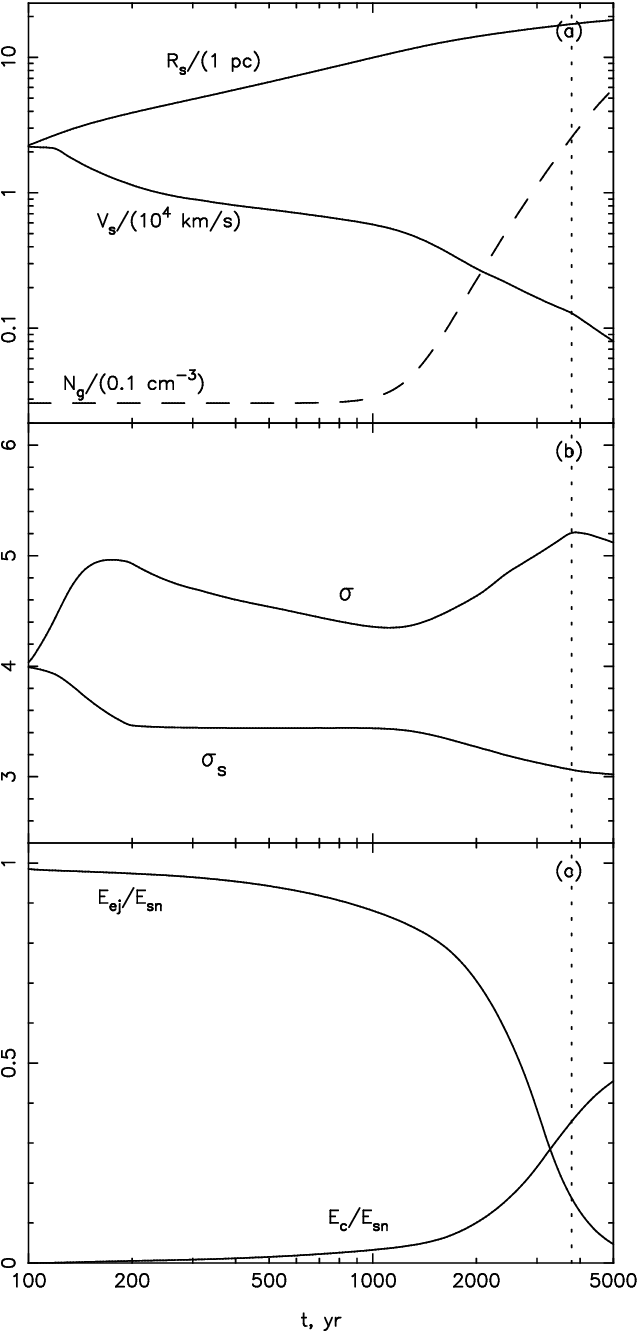}
\caption{Model parameters as a function of time: (a) Shock radius
  $R_\mathrm{s}$ and overall shock speed $V_\mathrm{s}$; (b) total shock
  ($\sigma$) and subshock ($\sigma_\mathrm{s}$) compression ratios; (c) ejecta
  ($E_\mathrm{ej}$) and CR ($E_\mathrm{c}$) energies in spherical symmetry.
  The vertical {\it dotted line} marks the current epoch of SNR
  evolution. The external gas density $N_{\mathrm{g}}=\rho_{\mathrm{g}} /
  m_\mathrm{p} = 1.4 N_{\mathrm{H}}$ is also shown in {\it panel (a)}.}
\label{f1}
\label{F:hydro_parameters}
\end{figure}
%------------------------------------------------------------------------------

To fit the observed synchrotron and \gr\ spectra (see below) we assume a proton
injection rate $\eta=10^{-3}$. This leads to a moderate nonlinear modification
of the shock which at the current age of $t_{\mathrm{sn}}=3745$~yrs 
has a total compression ratio $\sigma \approx 5.2$ and a subshock
compression ratio $\sigma_\mathrm{s} \approx 3.1$ (Fig.\,1b). All parameters
used and the resulting model properties are summarized in Table 1.

For its adopted density, the wind bubble contains only a small amount of gas
$M_{\mathrm {b}}\approx 0.3M_{\odot}$. Therefore the SN shock deposits only
about 20\% of the explosion energy during the initial $1000$~years of
propagation through the bubble, as seen in Fig.\,1. However, up to the current
epoch the SN shock has already swept up a considerable mass
$M_{\mathrm{sw}}\approx 25M_{\odot}$.
Therefore the ejecta have transformed about 85\% of their initial energy into
gas and CR energy (Fig.\,1c).
The acceleration process is therefore characterised by a high efficiency under
the assumption of spherical symmetry: at the current time about 35\%
of the explosion energy have been transferred to CRs, and the relative CR
energy content $E_{\mathrm{c}}$ continues to increase to a maximum of about
$0.55 E_{\mathrm{sn}}$
in the later phase (Fig.\,1c), when particles start to leave the source. 

Therefore, in absolute terms the CRs inside SNR \rxj\ already contain
\begin{equation}
E_{\mathrm{c}} \approx 0.35 E_{\mathrm{sn}} \approx 4.6 \times
10^{50}~\mathrm{erg}.
\end{equation}
% 

%-----------------------------------------------------------------------fig.2
\begin{figure} 
\centering 
\includegraphics[width=8.cm]{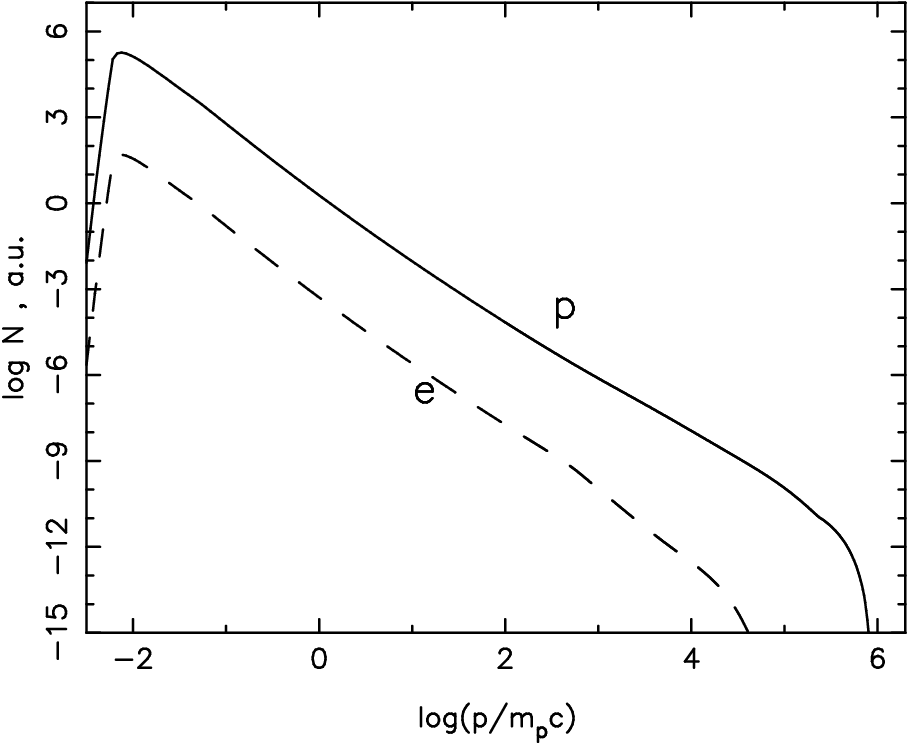}
\caption{The overall (volume-integrated) CR spectrum as a function of particle
momentum. {\it Solid and dashed lines} correspond to protons and electrons,
respectively.}
\label{f2}
\label{F:particle_spectra} 
\end{figure}
%------------------------------------------------------------------------------

The volume-integrated (or overall) CR spectrum 
 \begin{equation}
N(p,t)=16\pi^2p^2 \int_0^{\infty}dr r^2 f(r,p,t)  
\label{eq10}
\end{equation} 
has, for the case of protons, almost a pure power-law form $N\propto
p^{-\gamma}$ over a wide momentum range from $10^{-2}m_\mathrm{p}c$ up to the
cutoff momentum $p_\mathrm{max}\approx 6\times 10^5m_pc$, corresponding to a
cutoff energy of $~\approx 5.6 \times 10^{14}$~eV.  (see Fig.\,2).  This value
$p_\mathrm{max}$ is limited mainly by geometrical factors, which are the finite
size and speed of the shock, its deceleration and the adiabatic cooling effect
in the downstream region \citep{ber96}. Due to the shock modification the
power-law index slowly varies from $\gamma=2.4$ at
$p\,\,\raisebox{0.2em}{$<$}\!\!\!\!\!  \raisebox{-0.25em}{$\sim$}\,\,
m_\mathrm{p}c$ to $\gamma=1.9$ at $p\sim 100m_\mathrm{p}c$.  The shape of the
overall electron spectrum $N_\mathrm{e}(p)$ deviates from that of the proton
spectrum $N(p)$ at high momenta $p>p_\mathrm{l}\approx 350m_\mathrm{p}c$, as a
result of the synchrotron losses in the downstream region with a magnetic field
strength $B_\mathrm{d}\sim 100~\mu$G which is assumed uniform
($B_\mathrm{d}=B_2=\sigma B_0$). Therefore within the momentum range
$p_\mathrm{l}<p< p_\mathrm{max}^\mathrm{e}$ the electron spectrum is
considerably steeper $N_\mathrm{e}\propto p^{-3}$. 

Specifically, the synchrotron losses become important for electron momenta
greater than \citep{bkv02}
\begin{equation} 
\frac{p_\mathrm{l}}{m_\mathrm{p}c} \approx 
1.3 \left(\frac{10^8~\mbox{yr}}{t}\right)
\left(\frac{10~\mu\mbox{G}}{B_\mathrm{d}}\right)^2. 
\label{eq14}
\end{equation}
Substituting the SN age $t=3745$~yr
into this expression, we have $p_\mathrm{l}\approx 350m_\mathrm{p}c$, in
agreement with the numerical result
% (Fig.\,\ref{f2})} 
(see Fig.\,2).

%------------------------------------------------------------------------fig.3
\begin{figure*}
\centering
\includegraphics[width=0.85\textwidth]{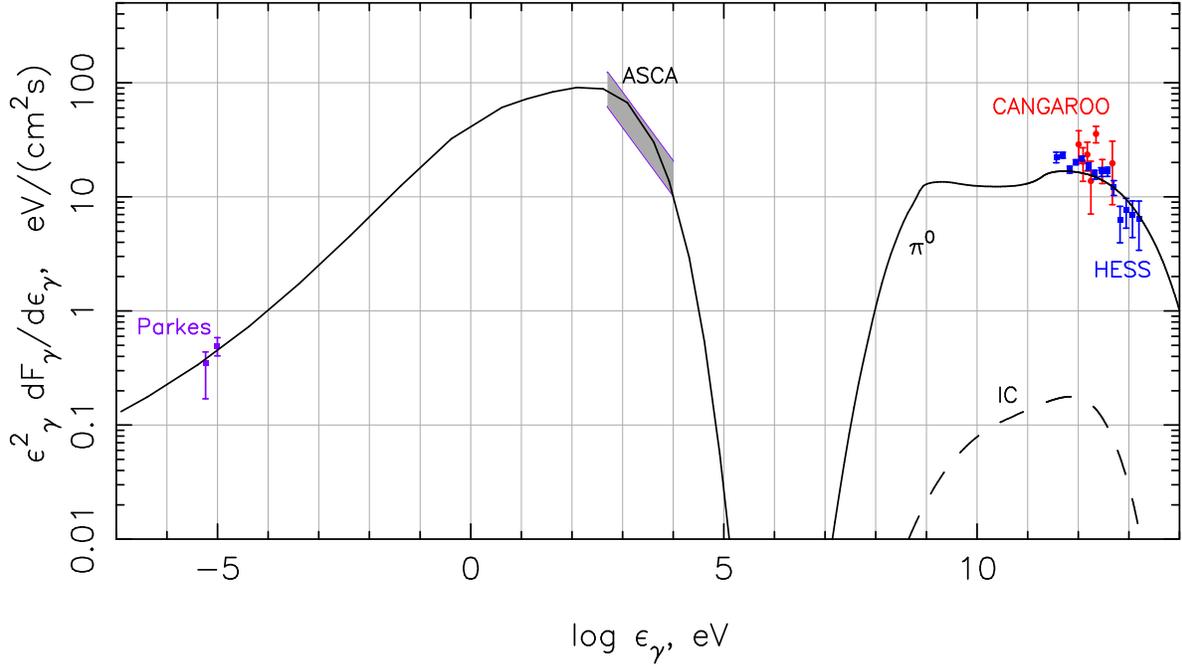}
\caption{Calculated broadband spectral energy density of \rxj\ , as function of
photon energy $\epsilon_{\gamma}$. In the $\gamma$-ray region the {\it solid
line} shows the $\pi^0-decay$~emission and the {\it dashed line} denotes the
inverse Compton (IC) emission. Radio fluxes were obtained with the {\it
Parkes} telescope \citep{duncang00}. For the X-ray synchrotron flux a lower
boundary is given by the sum of the {\it ASCA} fluxes from the brightest parts
of the SNR, as given by \citet{slane01}, scaled up to match the total remnant's
flux measured with {\it ROSAT} in the soft X-ray range \citep{aschen98}.  The
upper boundary comes from a re-analysis of the total {\it ASCA} data from the
remnant, as given in \cite{aha07a}. TeV data are from {\it CANGAROO}
\citep{enomoto06} and {\it H.E.S.S.}  \citep{aha07a}.}
\label{f3}
\label{F:broadband}
\end{figure*}
%------------------------------------------------------------------------------
%------------------------------------------------------------------------fig.4-
\begin{figure}
\centering
\includegraphics[width=8.cm]{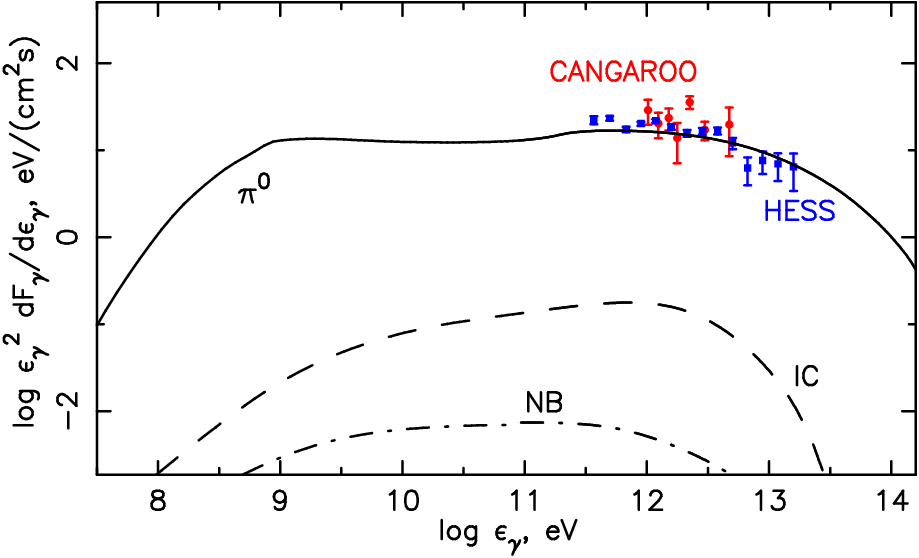}
\caption{Calculated nonthermal Bremsstrahlung {\it (NB, dash-dotted line)},
inverse Compton {\it (IC, dashed line)}, and $\pi^0$-decay {\it (solid line)}
$\gamma$-ray spectral energy distributions as functions of photon energy
$\epsilon_{\gamma}$ for the high-injection, high-field model. The
observed {\it H.E.S.S.} \citep{aha07a} and {\it CANGAROO} \citep{enomoto06}
$\gamma$-ray fluxes are shown as well.}
\label{f4} 
\label{F:gamma_spectra}
\end{figure}
%------------------------------------------------------------------------------

%------------------------------------------------------------------------fig.5
\begin{figure}
\centering
\includegraphics[width=7.5cm]{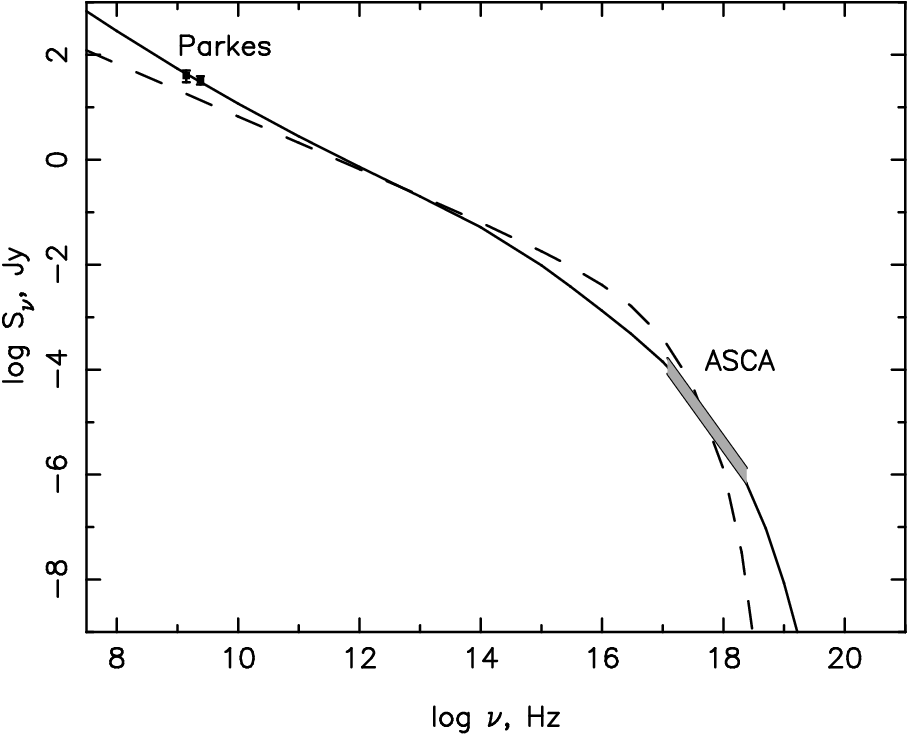}
\caption{Calculated synchrotron emission flux as a function of frequency
$\nu$. The {\it solid line} corresponds to the high-injection, high-field
model, the {\it dashed line} corresponds to a hypothetical very low injection,
and therefore unmodified low-field model. {\it Parkes} radio data
\citep{duncang00} and {\it ASCA} X-ray data \citep{slane01,aha07a} are shown
(see also the caption of Fig.\,\ref{f3}).}
\label{f5}
\label{F:synchrotron}
\end{figure}
%------------------------------------------------------------------------------

The maximum electron momentum can be roughly estimated by equating the
synchrotron loss time with the acceleration time. This gives
\citep[e.g.][]{bkv02}
\[
\frac{p_\mathrm{max}^\mathrm{e}}{m_\mathrm{p}c}
= 6.7\times 10^4 \left(\frac{V_\mathrm{s}}{10^3~\mbox{km/s}}\right)
\]
\begin{equation}
\hspace{1cm}\times
\sqrt{\frac{(\sigma-1)}{\sigma (1+ \sigma^2)}
  \left(\frac{10~\mu\mbox{G}}{B_0}\right)}. 
  \label{eq17}
\end{equation}

\noindent At the current epoch $V_\mathrm{s}\approx 1316$~km/s, which leads to
a maximum electron momentum $p_\mathrm{max}^\mathrm{e} \approx
10^4m_\mathrm{p}c$, in agreement with the numerical results (Fig.\,\ref{f2}).

As a result of the shock propagating through the wind shell (cf. Fig.\,1) the
SN shock speed decreases rather quickly during the period
$t>10^3$~yr. Therefore, during previous evolutionary phases the shock has
produced electron spectra with cutoff momenta $p_\mathrm{max}^\mathrm{e}$
larger than at the current epoch. Due to this factor the spatially integrated
electron spectrum has a relatively smooth cutoff (see Fig.\,2). Together with
the synchrotron cooling this gives a good fit of the observed X-ray spectrum
(see below).

The present-day parameters $B_\mathrm{d}= 104$~$\mu$G and $K_\mathrm{ep}\approx
3\times10^{-4}$ 
lead to good agreement between the calculated and the measured spectral energy
distribution of the synchrotron emission in the radio to X-ray ranges at the
present time (Fig.\,\ref{f3}). The steepening of the electron spectrum at high
energies due to synchrotron losses and the smooth cutoff of the overall
electron spectrum together naturally yield a fit to the X-ray data with their
soft spectrum. Such a smooth spectral behaviour is achieved in an assumed
upstream field of $20$~$\mu$G (which leads to the above downstream field
$B_\mathrm{d}$).

Fig.\,\ref{f3} also shows the calculated \gr\ spectral energy distributions due
to $\pi^0$-decay, IC emission, and nonthermal Bremsstrahlung, together with the
existing experimental data.

According to the calculation, the hadronic $\gamma$-ray production exceeds the
electron contribution by more than two orders of magnitude at all energies. In
detail these \gr\ spectra are shown in Fig.\,\ref{f4}. For energies
$\epsilon_{\gamma}=1-100$~GeV the \gr\ spectrum is close to
$dF_{\gamma}/d\epsilon_{\gamma}\propto \epsilon_{\gamma}^{-2}$, hardening from
$\epsilon_{\gamma}=0.1-1$~TeV, whereas starting from $\epsilon_{\gamma}\approx
1$~TeV it has a smooth extended cutoff despite the comparatively much sharper
cutoff of the proton energy spectrum, cf. Fig.\,\ref{f2}.

Note that the \gr\ cutoff energy $\epsilon_{\gamma}^{max}\approx
0.1p_\mathrm{max}c$ is sensitive to the magnetic field strength $B_\mathrm{d}$,
since the proton cutoff momentum has a dependence $p_\mathrm{max}\propto
R_\mathrm{s}V_\mathrm{s}B_\mathrm{d}$ \citep{ber96}.  It is clearly seen from
Fig.\,4 that the calculated spectrum fits the \hess\ measurements in an
acceptable way, at least up to $\approx 5$~TeV. However the four highest-energy
points tend to lie below the theoretical curve. This can be the result of
escape of the highest-energy protons during the deceleration of the shock in
the shell. See section \ref{SS:escape} for further details.

The hadronic dominance in VHE \gr\ emission which we predict here, could be
further investigated in the near future by the {\it Fermi} instrument in the
GeV region. Even though at such comparatively low \gr\
energies the \gr\ background from the diffuse Galactic \grs\ is quite
significant, especially for such a large, low-surface brightness source as
\rxj\ \citep{dav94}, {\it Fermi} should be able to detect the overall very high
\gr\ flux from \rxj. It is therefore to be expected that the {\it Fermi}
instrument will confirm our prediction that the spatially-integrated \gr\
spectral energy density at 1 GeV is only a factor $\approx 1.5$ {\it lower}
than at 1 TeV cf. Fig.\,4, as a result of the nonlinear modification of the
acceleration process. If the nonlinear modification in the wind bubble is less
strong than assumed here, then this difference in the spectral energy density
between 1 GeV and 1 TeV should be even smaller.

In Fig.\,5 we separately present the differential synchrotron spectrum,
produced at the current epoch. For comparison we also show a synchrotron
spectrum, which would correspond to a hypothetical acceleration scenario
in which the proton injection rate is taken so small ($\eta=10^{-5}$) that the
accelerated nuclear CRs do not produce any significant shock modification or
magnetic field amplification; the value $B_0=5$~$\mu$G is used in this
case. There are two small but distinct differences between the synchrotron
spectra that correspond to these two scenarios. The high-injection, high-field
scenario leads to a steep differential radio frequency spectrum $S_{\nu}\propto
\nu^{-\alpha}$ with power law index $\alpha\approx 0.7$, whereas for the
unmodified, low-field scenario $\alpha=0.5$. Unfortunately, the low quality of
the existing radio data does not allow us to distinguish these two scenarios in
the radio range, in order to conclude from the radio spectrum alone whether or
not we deal in the case of \rxj\ with efficient CR acceleration leading to a
significant shock modification and magnetic field amplification. The
essentially different behaviour of these two spectra at X-ray frequencies
around $\nu = 10^{18}$~Hz demonstrates on the other hand that in the case of
strong CR production and amplified magnetic field $B_\mathrm{d}\approx
100$~$\mu$G the spectrum $S_{\nu}(\nu)$ naturally exhibits a smooth cutoff
consistent with the experiment. In the simple, unmodified low-field case the
spectrum $S_{\nu}(\nu)$ has too sharp a cutoff to be consistent with the
experiment.

It is noted that the X-ray flux represented in Figs.\,3 and 5 comes from two
different analyses of the X-ray flux (see also section \ref{SS:broadband}): the
lower boundary was derived by summing up the {\it ASCA} fluxes from the
brightest parts of the SNR \citep{slane01} and scaling the result up to
match the total SNR's flux as measured with ROSAT \citep{aschen98}.  The
upper boundary -- a factor of two higher -- comes from a re-analysis of the
total {\it ASCA} data from the remnant, as given in \citet{aha07a}.  

\begin{figure}
\centering
\includegraphics[width=7.5cm]{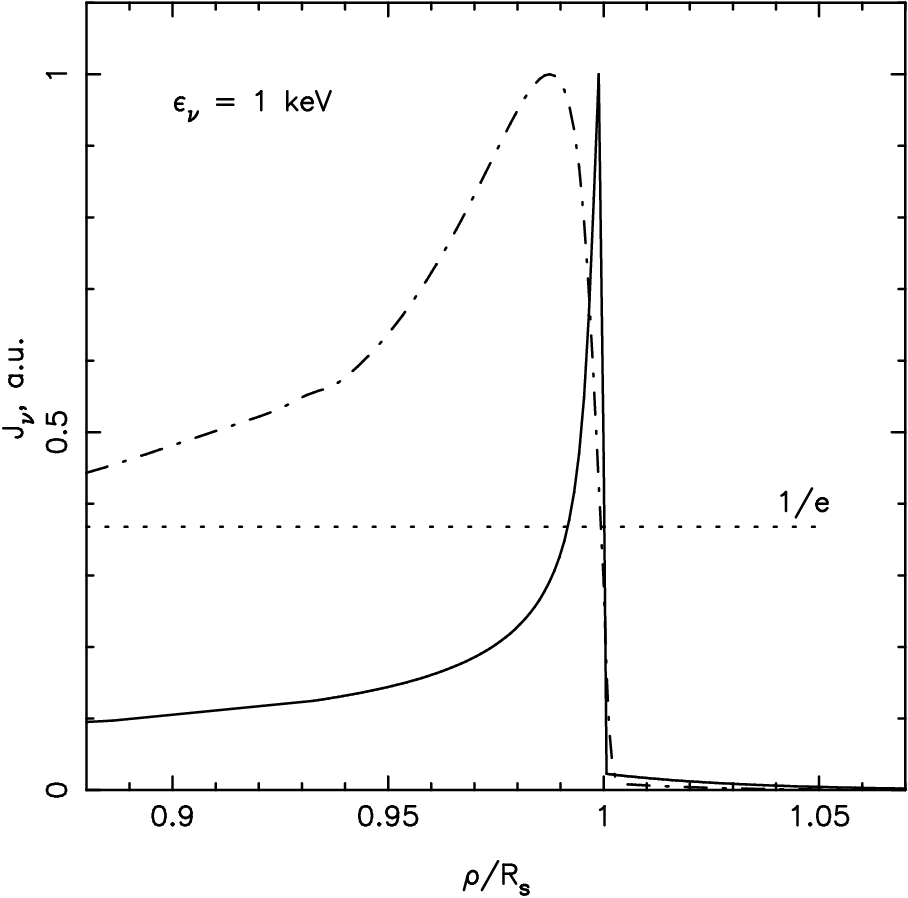}
\caption{The projected radial profile of the X-ray synchrotron emission for
$\epsilon_\mathrm{\nu}=1\,\mathrm{keV}$. The abscissa is normalized to the
shock radius and corresponds therefore to an angular scale.  The {\it solid
line} corresponds to the high-injection, high-field model, the {\it dash-dotted
line} corresponds to a hypothetical low-injection, low-field model.}
\label{f6}
\label{F:xrayradial}
\end{figure}

\begin{figure}
\centering
\includegraphics[width=7.5cm]{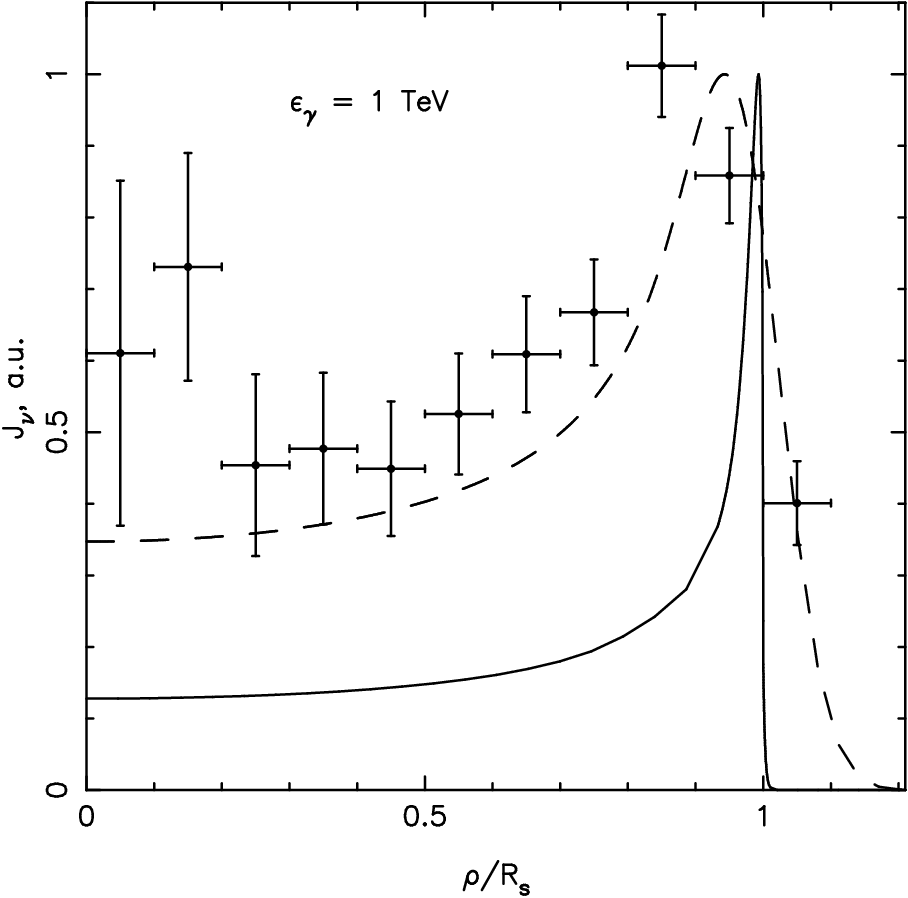}
\caption{The \gr\ emissivities for the \gr\ energy
$\epsilon_{\gamma}=1\,\mathrm{TeV}$ as a function of projected, normalized
radial distance $\rho/R_\mathrm{s}$. The calculated radial profile is
represented by the {\it solid line}. Data points are from the Northern part of
\rxj\ as measured with {\it H.E.S.S.} \citep{aha07a}, with an analysis point
spread function of Gaussian width $0.06^\circ$. The {\it dashed line}
represents the calculated profile convolved with the same point spread
function.}
\label{f7}
\label{F:gammaradial}
\end{figure}

The properties of small-scale structures of \rxj\ seen in X-rays furnish even
stronger evidence that the magnetic field inside the SNR is indeed considerably
amplified. As in the case of other young SNRs (e.g. SN~1006, Cassiopeia~A,
Tycho's SNR) {\it Chandra} shows very fine filamentary structures in nonthermal
X-rays in the very outer part of the remnant. The thinnest filament detected by
\citet{bamba05} has an angular thickness $\Delta \psi\approx 38''$ in the
radial profile of the X-ray emission in the 2 -- 10~keV range. In order to find
out whether this type of structure is consistent with our theory we present in
Fig.\,6 the projected radial profile
\begin{equation}
J(\epsilon,\rho)\propto 
\int dx q(\epsilon,r=\sqrt{\rho^2+x^2},x),
\end{equation}
calculated for the X-ray energy $\epsilon_{\nu}=1$~keV. The abscissa in
Fig.\,\ref{F:xrayradial} (and correspondingly also in
Fig.\,\ref{F:gammaradial}) is scaled to the radius of the remnant,
i.e. to 17.5\,pc, as derived from the angular radius of the remnant of $\sim
1^{\circ}$.  Here $q(\epsilon, r)$ is the luminosity in the nonthermal emission
with photon energy $\epsilon$. The integration is performed along the line of
sight. One can see that the theory predicts the peak of the emission just
behind the shock front with a thickness $\Delta \rho/R_\mathrm{s}\approx
10^{-2}$ which corresponds to an angular width $\Delta \psi\approx 36''$. It
corresponds reasonably well to the {\it Chandra} observation. At the same time
Fig.\,6 also shows that the unmodified low-field solution cannot explain the
filament structure seen with {\it Chandra}.

%-------------------------------------------------------------------fig.6
%-------------------------------------------------------------------------

It has been already demonstrated for other young SNRs \citep{bkv03,bv04a} that
the measured width of the projected radial profile of the nonthermal X-ray
emission gives the possibility to determine the internal magnetic field
strength according to Eq.(1). 

Substituting into this equation $l_2=L/7=8.2\times 10^{16}$~cm, $\sigma=5.2$,
$V_\mathrm{s}=1316$~km/s, and $\nu =3\times 10^{17}$~Hz (i.e. X-ray energy
$\epsilon_{\nu}\approx 1$~keV), we obtain $B_\mathrm{d} \approx 139$~$\mu$G,
which agrees within 30\% with the value $B_\mathrm{d}=104$~$\mu$G used in our
spectrum calculation. Such a difference in $B_\mathrm{d}$ corresponds to the
uncertainty in the field determination in the other objects analysed up to
now.

The magnetic field amplification is driven by the gradient of the (nuclear) CR
pressure upstream of the outer shock, and we can check whether \rxj\ belongs to
the class of objects that fulfil Eq.\,2.  In the present case of $d=1$~kpc we
have $P_\mathrm{c}\approx 0.15\rho_0 V_\mathrm{s}^2$, where
$\rho_0=N_\mathrm{g}(R_\mathrm{s})m_\mathrm{p}$ is the ambient gas density at
the current shock front position. Substituting
$N_\mathrm{g}(R_\mathrm{s})=0.24$~cm$^{-3}$ and $B_0=20$~$\mu$G we have
$B_0^2/(8\pi P_\mathrm{c}) \approx 6.5\times 10^{-3}$, in rather good agreement
with the average number in Eq.(2).

The line-of-sight integrated \gr\ emission profile as a
function of projected radius $\rho$ is calculated for
$\epsilon_\mathrm{\gamma}=1$~TeV and is presented in Fig.\,7.  Due to the large
radial gradient of the gas and CR distributions inside the SNR, the
theoretically predicted three-dimensional radial emissivity profile of
TeV-emission is concentrated within a thin shell of width $\Delta r\sim
0.01R_s$. As a result of the projection effect, the two-dimensional width is
by a factor of seven larger than the width of the three-dimensional profile,
i.e. $\Delta \rho\approx 0.1R_{\mathrm {s}}$ (solid lines in
Fig.\,\ref{F:gammaradial}). Since the {\it H.E.S.S.} instrument in
addition has a finite angular resolution we present as dashed lines in Fig.\,7
also the modified radial profile convolved with a Gaussian point spread
function with a $\sigma_{\rho}=\Delta \rho = 0.06R_\mathrm{s}$, corresponding
to an angular resolution of $\sigma_{\psi} = 0.06^{\circ}$.

%-------------------------------------------------------------------fig.7
%-------------------------------------------------------------------------

Fig.\,7 shows that the expected radial profile of the TeV-emission, after
taking into account the instrumental angular resolution, is much broader than
the intrinsic projected profile and is characterised by a minimum-to-maximum
intensity ratio $J^\mathrm{min}_{\gamma}/J^\mathrm{max}_{\gamma} \approx
0.35$. The radial profile measured by {\it H.E.S.S.} compares reasonably well
with the theoretical prediction if we ignore the two data points in the central
region.  This might be justified because a central component (e.g. a PWN)
cannot be fully excluded, as was also argued in \citet{aha07a}.

%-------------------------------------------------------------------fig.8
\begin{figure}
\centering
\includegraphics[width=7.5cm]{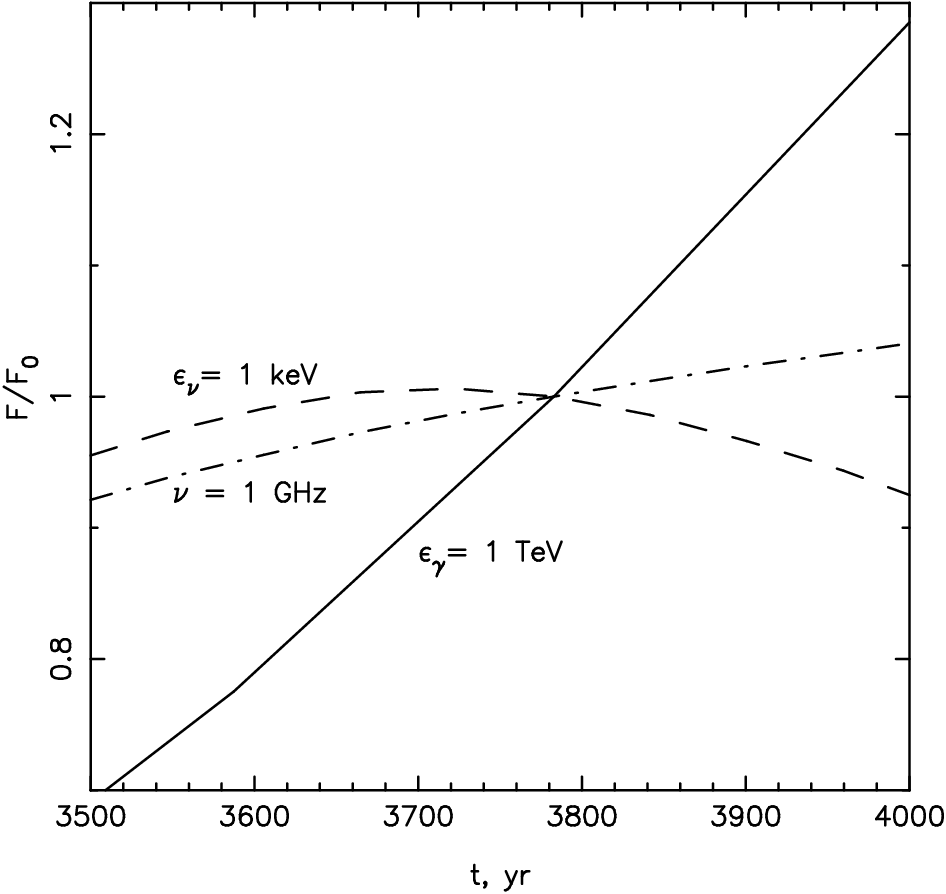}
\caption{The time dependence of the fluxes of the synchrotron emission at
frequency $\nu=1\,\mathrm{GHz}$ {\it (dash-dotted line)}, synchrotron X-ray
emission with energy $\epsilon_{\nu}=1\,\mathrm{keV}$ {\it (dashed line)} and
TeV-energy \gr\ {\it (solid line)}. The fluxes are normalised to their values
$F_0$ at the current epoch.}
\label{f8}
\label{F:secular}
\end{figure}
%-------------------------------------------------------------------------
%

To illustrate the expected time evolution of the nonthermal emission we give in
Fig.\,8 the fluxes of radio emission at the frequency $\nu=1$~GHz and X-ray
emission at the energy $\epsilon_{\nu}=1$~keV, as well as the TeV flux, all as
a function of time. Since the CR energy content still increases, the radio
emission increases during the next centuries. However, due to the substantial
shock deceleration the nonthermal X-ray emission is expected to decrease with
time at a rate of about 0.04\%/yr. The TeV \gr\ emission, on the other hand,
will increase at a rate of 0.14 \%/yr due to the continuing increase of the
ambient gas density. Nevertheless, a change of the order of one percent in 7
years is probably very difficult to measure. Therefore we can not put much
practical weight on these expectations regarding the secular variation of
nonthermal fluxes.

\subsection{Escape}
\label{SS:escape}
 
The decrease of the shock speed $V_\mathrm{s}$ during SNR evolution diminishes
the maximum energy to which particles can be accelerated during any given
phase. This has the well-known consequence that particles already accelerated
to a higher maximum energy at earlier times can now begin to diffusively escape
from the interior of the remnant, even without any change of the scattering
mean free path, as it was predicted analytically \citep{bek88} and confirmed
numerically \citep{byk96}.

Compared with this escape -- which is relatively slow in our model because the
diffusion of CR particles with momenta $p>p_\mathrm{m}(t)$ higher than the
current cutoff momentum $p_\mathrm{m}(t)$ is still described by Bohm diffusion
-- the actual CR escape is presumably much faster. Since the CRs with
$p>p_\mathrm{m}(t)$ develop a much smaller spatial gradient than those with
momenta $p<p_\mathrm{m}(t)$ (which are still efficiently accelerated by the
shock), their production of chaotic magnetic fields whose scales are
approximately equal to their gyroradii goes down quickly with time. As a
consequence their diffusion coefficient increases considerably so that they
leave the SNR much more rapidly than through Bohm diffusion, and then these CRs
contribute {\tiny much} less to the \gr\ flux of the source than calculated
above. The \gr\ spectrum corresponding to the particles
that remain confined in the SNR should therefore drop off faster with energy
than given by the solid line in Fig.4. However, the \hess\ data at the highest
\gr\ energies are not suggesting that this effect is very significant in the
present case.

Secondly, the field amplification is probably stronger during earlier phases of
the remnant evolution, when $\rho_0 V_\mathrm{s}^2$ is larger than at later
phases \citep{belll}. Our calculation has assumed the effective field strength
to be constant. Correcting for this, the overall nuclear particle spectrum
released into the Interstellar Medium by SNRs presumably forms the Galactic CR
spectrum up to an energy of $10^{17}$~eV \citep{bv07}. That protons alone
should be able to reach energies $\gsim 10^{15}$~eV was shown for the case of a
representative type Ia SN by \cite{bv04b}. This is likely to be true also for
\rxj\ and, by implication, also for its twin RX~J1713.7-3946, even though their
environmental conditions are drastically different from those of a type Ia SN.

We add here that the above maximum energies of energetic nuclei are
calculated under the assumption of Bohm diffusion. To this extent they are
upper limits, given the strength of the corresponding amplified magnetic
field.

Except in special cases one should therefore expect that escape sets in when
the source becomes older. \citet{PtuskinZirak2003,PtuskinZirak2005} have in
addition argued that at late times the damping of the scattering magnetic
fluctuations should increase the scattering mean free path. The decrease of the
effective field strength on the one hand, and of the scattering strength
(through the slowing-down of the shock and wave damping) on the other, go in
the same direction to lower the cutoff energy of the accelerating
population. The relative importance of these effects probably depends on the
details of the source, in particular its evolutionary phase.

\subsection{A rough estimate for the thermal X-ray emission}

Using the results for the dynamical evolution of the system one can also 
  attempt to estimate the thermal X-ray emission. As argued in section 3.1.4
this is only possible in an approximate way, even if the
ejecta emission is disregarded with the argument that the ejected mass in
the present phase is small compared to the swept-up mass. In fact, looking at
Fig. 1a it can be seen that the SNR is well past the sweep-up phase and has
entered a quasi-Sedov phase in the stellar wind shell, i.e. a roughly
self-similar evolutionary phase, modified by strong particle acceleration
relative to a purely gas dynamic evolution.

The approximation used is the following. The bubble case is compared with a SNR
in a uniform medium in the classical Sedov phase without any CR acceleration,
making four assumptions (i) the total hydrodynamic explosion energy is the same
in both cases (ii) the downstream unmodified temperature $T_\mathrm{s}$,
determining the overall shock velocity and the thermal emission, is the same,
(iii) the present gas density upstream of the shock is the same, and (iv) the
two objects are at the same distance of 1 kpc. Then the results of
\citet{hsc83} for the thermal X-ray flux from a classical Sedov SNR are used,
employing the emission measure of the bubble remnant instead of that of the
classical Sedov remnant with the same five parameters above. This means that
the X-ray emissivity of the remnant is reduced by the ratio $R_\mathrm{em}
=EM_\mathrm{b}/EM_\mathrm{S}$ of the emission measure $EM_\mathrm{b}$ for the
bubble solution to the emission measure for the classical Sedov solution
$EM_\mathrm{S}$ which corresponds to a uniform ambient gas density
$N_\mathrm{g}(R_\mathrm{s})$.

The thermally relevant gas temperature in the presence of particle acceleration
is the temperature $T_\mathrm{sub}$ downstream of the gas subshock. It
corresponds to the temperature downstream of a shock that is nonlinearly
modified by the internal energy and the pressure of the accelerated particles.

The CR-modified shock, i.e. the precursor-subshock system, is approximated by a
plane parallel structure, which implies that the precursor size is small
compared to the SNR shock radius. This is generally the case. Then
conservation of mass, momentum and energy fluxes permits in a straightforward
way to calculate the ratio between the actual downstream gas temperature
$T_\mathrm{sub}$ and the downstream gas temperature $T_\mathrm{s}$ that would
obtain without particle acceleration:

\begin{equation}
  \frac{T_\mathrm{sub}}{T_\mathrm{s}}=
  \frac{[(\sigma_\mathrm{s} -1)(\gamma+1)+2](\gamma+1)^2}
{4 \gamma (\gamma -1) \sigma^2}.
\end{equation}

Here $\sigma$ and $\sigma_\mathrm{s}$ are the total and the subshock
compression ratio, respectively, as given in Fig.1. The specific heat ratio
$\gamma$ for the nonrelativistic thermal gas may be
taken as $\gamma = 5/3$.
For the present phase of \rxj\ , i.e. $\sigma \approx 5.2$ and
$\sigma_\mathrm{s} \approx 3.1$, one then obtains
$T_\mathrm{sub}\approx 0.45T_\mathrm{s}$. 

\citet{hsc83} give the gas temperature $T_\mathrm{s}$ by
\begin{equation}
T_\mathrm{s}=10^7(V_\mathrm{s}/839~
      \mathrm{km~ s}^{-1})^2 \mathrm{K}
\end{equation}
in terms of the overall shock velocity $V_\mathrm{s}$, assuming a highly
  ionized system. For \rxj\ the latter quantity is given in our Fig.1a. Using
the present value $V_\mathrm{s}=1316$~km/sec, the foregoing equations result in
$T_\mathrm{s}\approx 2.5 \times 10^7$~K and $T_\mathrm{sub} \approx 1.1\times
10^7$~K, corresponding to a thermal energy of about 1 keV.

For the emission measure EM we have $EM=\int_0^{R_\mathrm{s}}dr r^2
N_\mathrm{g}^2(r)$, if we disregard irrelevant numerical factors.  In order to
calculate $R_\mathrm{em}$ for the CR-modified shock the overall shock is
approximated by a shock in a thermal gas with an adiabatic index
$\gamma_\mathrm{cr}$ such that $(\gamma_\mathrm{cr} +1)/(\gamma_\mathrm{cr} -1)
= \sigma$, where $\sigma$ denotes, as before, the total compression ratio of
the CR-modified shock. The corresponding solution for the gas dynamic
quantities is approximated as being self-similar cf. \citet{sedov}.  If the
ambient density profile has a radial dependence in the form of a power law
$N_\mathrm{g}(r)=N_0(r/R_0)^{\beta}$, then the self-similar density
distribution in the SNR, downstream of the shock, can be further approximated
by a power-law profile, with the same swept-up mass as the full solution. It
has then the form $N_\mathrm{g}(r)=\sigma N_0(r/R_\mathrm{s})^{\beta'}$ with
$\beta'= 3(\sigma -1)+ \beta\sigma $, where
$N_0=N_\mathrm{g}(R_\mathrm{s})$. Such a density profile gives the emission
measure $EM=N_0^2R_\mathrm{s}^3\sigma^2/(2\beta'+3)$.

The classical gas dynamic Sedov solution for a SN explosion into a uniform
medium has $\beta = 0$ and $\sigma = 4$. Using the same values for $N_0$ for
both density profiles leads to $R_\mathrm{s}\approx 12.51$~pc for the Sedov
case and to $R_\mathrm{em} \approx 0.63$ for the bubble case $\beta=12$ and
$\sigma=5.2$. The swept-up mass in the Sedov case is $M_\mathrm{S} \approx
50.3~M_{\odot}$.

Using the differential thermal X-ray model spectra $dF/d\epsilon\approx
10^{-4}$~photons/(keV cm$^2$s)~from \citet{hsc83} (see their Fig. 2) for their
$\lg T_s=7.75$ and $\lg T_s=7.25$,
$\eta = N_\mathrm{H}^2 E_\mathrm{sn}=10^{49}$~erg cm$^{-6}$, scaling it
according to $dF/d\epsilon\propto \eta$ with a factor of $\approx 3.8$ for
$\eta \approx 3.8\times 10^{49}$~erg cm$^{-6}$, multiplying it by the factor
$\theta^2 [E_\mathrm{sn}/(10^{51}\mathrm{erg})]^{-1/2} \approx 0.65 \times
10^4$, as required, where $\theta \approx 86$~arcmin for the angular size of
the classical Sedov remnant corresponding to the above parameters
\citep{hsc83}, and multiplying it finally also by the factor $R_\mathrm{em}$,
results in a thermal spectral energy density $\epsilon^2 dF/d\epsilon \approx
1560$~eV cm$^{-2}$ sec$^{-1}$ for $\epsilon =1$~keV.

This must be compared with the observed nonthermal X-ray energy flux at 1 keV
(see Fig.3) $\epsilon^2 dF/d\epsilon \approx 100$~eV cm$^{-2}$
sec$^{-1}$. Therefore, at 1~keV, the thermal flux, calculated in this form,
comes out to be larger than the nonthermal flux by a factor of about 16.

We note however that the above X-ray energy flux has to be considered as a
rough upper limit estimate because of the following reasons:

First of all, the SN shock into the bubble and shell interacts with a strongly
rising gas density profile. Yet for the estimate the peak value of the gas
density was used.

Moreover,in the case of a modified shock the actual gas temperature
$T_\mathrm{sub}$ is at least by a factor of 2 lower than the temperature
$T_\mathrm{s}$ used in our estimate. From a general point of view the actual
thermal X-ray emission is expected to be lower due to lower temperature. A
rough correction can be done in the following way. The shock radius
corresponding to the Sedov solution depends on the relevent parameters
according to the relation $R_s\propto (E_{SN}/(T_s
N_\mathrm{H}))^{1/3}$. According to this relation the shock with the same
radius but with two times lower postshock temperature corresponds to a two
times lower explosion energy. Such a shock is characterised by the parameter
value $\eta \approx 1.9\times 10^{49}$~erg cm$^{-6}$ that gives a 1~keV thermal
X-ray flux which is lower by a factor of two compared with the above
consideration.

Third, as it was already mentioned, the calculation in section 4 was performed
for a rather moderate value of the injection rate, which leads to a moderate
shock modification. Higher injection rates can not be excluded given the
present data which can lead to a considerably higher shock modification with
significantly lower gas energy and gas temperature. For a comparatively extreme
position, see the recent paper by \citet{damg08}.

Fourth, for a fixed \gr\ flux $F_{\gamma}\propto N_H E_{SN}/d^2$~the required
gas number density scales as $N_H\propto d^2$ with source distance $d$.
Consequently a decrease of $d$ requires a decrease of the gas density.  The
thermal X-ray luminosity is quite sensitive to the gas density and to the total
gas energy because roughly $dF/d\epsilon\propto N_H^2 E_{SN}$. A somewhat
smaller distance than the $d=1$~kpc, assumed here, can not be excluded.

Finally, the bubble solution developed in this paper has quite a different
temperature profile than any ``equivalent'' classical Sedov solution used to
estimate the thermal emission. This arises from the approximate uniformity of
the total pressure in the SNR interior which roughly implies $T_\mathrm{sub}
\propto N_\mathrm{H}^{-1}$. The density profiles are indeed quite different in
both cases.

As a result, we believe that the uncertainties regarding the thermal X-ray
emission are quite large. This situation also implies some systematic error in
the overall model presented, although this can hardly change its key
properties. In our opinion the uncertainty in the thermal emission has to be
resolved by future work that extends the efforts of \citet{hsc83} to more
general circumstellar density profiles and ejected masses for core-collapse
supernovae.

\section{Conclusions}

We have argued in the last section that a solution behind the Vela SNR,
involving the core collapse of a massive star in its own wind bubble, describes
the available data reasonably well. The physical characteristics of this
solution are very similar to those of the well-known SNR RX~J1713.7-3946. In
this sense \rxj\ and SNR RX~J1713.7-3946 are twins.

The calculated \gr\ spectrum has a smooth cutoff at higher energies.  The
presently observable lower limit for the magnetic field amplification --
in terms of thin shock filaments in hard X-rays -- is consistent with the one
deduced from a theoretical fit to the observed synchrotron spectra. If
anything, the spectrally deduced field amplification
is lower than the filament-deduced field amplification. Overall, the
field amplification is also consistent with the semi-empirical
relation Eq.(2) between the CR pressure $P_\mathrm{c}$ and the magnetic
pressure, where $P_{\mathrm{c}}$ stems from the proton injection rate required
to fit the observed \gr\ emission. We conclude from these consistency arguments
that the observational lack of a detailed radio synchrotron spectrum does not
preclude the determination of a consistent amplified field.

The magnetic field amplification results in a significant depression of the
density of ultra-high energy electrons and reduces the IC and NB contributions
of these electrons to less than one percent of the $\pi^0$-decay \gr\ emission.

A remaining uncertainty is connected with the thermal emission properties. For
the wind bubble configuration this may indeed not be too critical,
since gas heating should occur primarily at the subshock
alone. In addition, the amount of swept-up mass is rather small compared to a
classical Sedov remnant in a uniform circumstellar medium of equal
present preshock density.

Apart from this uncertainty the main result is the hadronic dominance in
the \gr\ emission spectrum. As a consequence of the nonlinear modification of
the shock the spatially integrated \gr\ spectral energy distribution at 1 GeV
is predicted to be at best a factor $1.5$~ lower than at 1 TeV.

\begin{acknowledgements}
  This work has been supported in part by the Russian Foundation for Basic
  Research (grants 06-02-96008, 07-02-0221). The authors thank V.S. Ptuskin and
  V.N. Zirakashvili and the anonymous referee for discussions on the
  thermal emission properties. EGB acknowledges the hospitality of the
  Max-Planck-Institut f\"ur Kernphysik, where part of this work was carried
  out.
\end{acknowledgements}

\end{document}